\documentclass[showpacs,preprint]{revtex4}
\usepackage{mathrsfs}
\usepackage{graphicx}
\usepackage{subfigure}
\usepackage{amssymb}
\usepackage{amsmath}
\usepackage{bm}
\usepackage{latexsym}
\usepackage{hyperref}

\begin{document}

\begin{titlepage}
\title
{Non-Markovian disentanglement dynamics of two-qubit system}
\author
{Xiufeng Cao\footnote{Email: cxf@sjtu.edu.cn}, Hang Zheng}
\address{Department of Physics, Shanghai Jiao Tong University,
Shanghai 200240, People's Republic of China}

\begin{abstract}
We investigated the disentanglement dynamics of two-qubit system in
Non-Markovian approach. We showed that only the couple strength with
the environment near to or less than fine-structure constant 1/137,
entanglement appear exponential decay for a certain class of
two-qubit entangled state. While the coupling between qubit and the
environment is much larger, system always appears the sudden-death
of entanglement even in the vacuum environment.
\\


\end{abstract}

 \maketitle

\end{titlepage}

\section{Introduction}

A multipartite quantum system, in addition to local quantum coherence that
exists within each of subsystems, may have nonlocal or distributed quantum
coherence that exists among several distinct subsystems. This property is
entangle, which is superposition of the internal states of the systems and
cannot be separated into product states of the individual subsystem. It is
recognized as entirely quantum-mechanical effect and have played a crucial
role in practical application ranging from quantum information\cite{A3,A4},
cryptography\cite{A5} and quantum computation\cite{A6,A7}, to atomic and
molecular spectroscopy\cite{A8,A9}.

Recent, many groups were able to prepare entangled states in a variety of
physical systems and experimental setups, demonstrating an impressive
ability to manipulate and detect them efficiently.\cite%
{B1,AA1,AA2,AA3,AA4,AA5,AA6,AA7} In particular, Almeida \textit{et al}.\cite%
{B1} showed that, using an all-optical experimental setup, even when the
environment-induced decay of each system is asymptotic, quantum entanglement
may appear "entanglement sudden death", called ESD\cite{C3}, that is,
entanglement terminates completely after a finite interval, without a
smoothly diminishing long-time tail.

As we known, a major obstacle for the controlled entanglement of more and
more subsystems remains with the capacity of achieving perfect screening of
the system from the environment. After some time, the unavoidable residual
interaction with the environment induces mixing of the system state, and
thus the emergence of classical correlations at the expense of quantum
entanglement. Hence, we face the high relevant task of understanding the
sources of entanglement decay, what implies the identification of the
associated time scales. In addition to be enlightened by the experimental
discovery of ESD, a large number of theoretical literature have investigated
the disentanglement dynamics.\cite{C1,C2,C3,C4} Z. Ficek and R. Tana\'{s}
\cite{C1} propose the review with an overview of the mathematical apparatus
necessary for describing the interaction of atoms with the electromagnetic
field. Present the master equation technique and describe a more general
formalism based on the quantum jump approach. F. Mintert \textit{et al.}\cite%
{C2} start with a short recollection of environment models adapted for
decoherence process in a typical quantum optical context under the
assumption of complete positivity and Markovian dynamics in the Lindblad
form. Yu and Eberly \cite{C3,C4} showed that the dynamics of the quantum
entanglement between two qubits interacting independently with either
quantum noise or classical noise displays a completely different behavior
from the dynamics of the local coherence. Instead of the exponential decay
in time of the local coherence, quantum entanglement may disappear within a
finite time in the dynamical evolution. The "entanglement sudden death" has
been experimentally demonstrated by Almeida. However, it is surprising that
few if any fundamental treatment exist of decoherence that include the
dynamics of disentanglement on better than an Markovian approximation or
phenomenological. Although, the use of the Markovian approximation is
justified in a large variety of quantum optical experiments where
entanglement has been produced, one should notice that Non-Markovian effects
are important in the description of some condensed-matter system\cite{D1},
such as the quantum dot qubit(s) system. Therefore, a Non-Markovian effects
of the decoherence in any viable realization of qubits is desirable.

In this paper we examined the disentanglement dynamics of two entangled
qubits due to spontaneous emission, where the interaction with the
environment without rotating-wave approximation and the treatment process
without Markovian approximation. It is found that disentanglement always
take only a finite-time to be completed, called "entanglement sudden death",
when the coupling between qubit and environment is strong. While the
coupling with dissipation environments is weak to fine structure constant $%
1/137$, the disentanglement change from exponential decay to entanglement
sudden-death with the increasing of the portion of the double excitation
component in the initial entangled state. We describe the sudden-death time
of entanglement or the realized lifetime of the given two-qubit entanglement
system through the measurable parameters: coupling constant with the
enviornment $\alpha ,$ energy splitting $\Delta $ and cut-off frequency $%
\omega _{c}$. If we consider entanglements as the central resource of most
types of quantum information processing, it is the most relevant question in
entanglement experiment under the environment-induced mixing.

The paper is organized as follows: In sec. \ref{model} we introduce the
Hamiltonian without rotating-wave approximation in the two-qubit environment
interaction and solve it in terms of Non-Markovian treatment. The dependence
of the concurrence on the different initial condition and the coupling
strength to the dissipation environment, are discussed in sec. \ref%
{discussion}. Finally, the conclusion is given in sec. \ref{conclusion}.

\section{The model and theory}

\label{model}

This paper is concerned primarily with two two-level systems, since it is
generally believed that entanglement of only two microscopic quantum systems
(qubits, atoms) is essential to implement quantum protocols such as quantum
computation. We consider two two-level subsystem A, B and assume that each
subsystem interacts independently with the environment, a well justified
assumption wherever the particles composing your system are sufficiently
separated from each other, and therefore, no collective environment effects
must be taken into account. In non-rotating wave form such a model may be
formulated to the following total Hamiltonian (set $\hbar =1$):
\begin{equation}
H(t)=H_{qu}+H_{env}+H_{int},
\end{equation}%
with%
\begin{equation}
H_{qu}=-\frac{1}{2}\Delta _{A}\sigma _{z}^{A}-\frac{1}{2}\Delta _{B}\sigma
_{z}^{B},
\end{equation}%
\noindent
\begin{equation}
H_{env}=\sum_{k}\omega _{k}a_{k}^{+}a_{k}+\sum_{k}\nu _{k}b_{k}^{+}b_{k},
\end{equation}%
\noindent
\begin{equation}
H_{int}=\frac{1}{2}\sum_{k}g_{k}(a_{k}^{+}+a_{k})\sigma _{x}^{A}+\frac{1}{2}%
\sum_{k}f_{k}(b_{k}^{+}+b_{k})\sigma _{x}^{B},
\end{equation}%
where the Hamiltonian of the two qubits $H_{qu}$, the two independently
environments $H_{env}$, the interaction $H_{int}$. Here \noindent \noindent $%
\sigma _{i}$ $(i=x,y,z)$ denotes the usually Pauli spin matrices, $\Delta
_{A}$ ($\Delta _{B}$) describes the energy splitting in the A (B) qubit. $%
a_{k}^{+}$ ($b_{k}^{+}$)$,$ $a_{k}$ ($b_{k}$) and $\omega _{k}$ ($\nu _{k}$)
are the creation, annihilation operator and energy with wave vector $k$ in
the A (B) qubit environment. $g_{k}$ and $f_{k}$ are the qubit-environment
coupling strength. Yu and Eberly \textit{etc.}\cite{C3,C4} has employed the
similar model, but the rotating-wave approximation is valid. Two
environments are completely defined by the spectral density:%
\begin{equation}
J(\omega )=\sum_{k}g_{k}^{_{2}}\delta (\omega -\omega _{k}).
\end{equation}%
We consider the Ohmic bath $J(\omega )=2\alpha \omega \theta (\omega
_{c}-\omega )$ in this work, where $\alpha $ is the dimensionless coupling
constant and $\theta (x)$ is the usual step function.

In order to simplify the non rotating-wave term, we apply a canonical
transformation, $H^{\prime }=\exp (s)H\exp (-s)$ with the generator\cite{E2}:%
\begin{equation}
S=\sum_{k}\frac{g_{k}}{2\omega _{k}}\xi _{k}^{A}(a_{k}^{+}-a_{k})\sigma
_{x}^{A}+\sum_{k}\frac{f_{k}}{2\nu _{k}}\xi _{k}^{B}(b_{k}^{+}-b_{k})\sigma
_{x}^{B}.
\end{equation}%
Then decompose the transformed Hamiltonian $H^{\prime }$ into three parts:%
\begin{equation}
H^{^{\prime }}=H_{0}^{^{\prime }}+H_{1}^{^{\prime }}+H_{2}^{^{\prime }},
\end{equation}%
where the three parts include the analogous form for A and B qubit,%
\begin{equation}
H_{0}^{^{\prime }}=H_{0A}^{^{\prime }}+H_{0B}^{^{\prime }}
\end{equation}%
with%
\begin{equation}
H_{0A}^{^{\prime }}=-\frac{1}{2}\eta ^{A}\Delta _{A}\sigma
_{z}^{A}+\sum_{k}\omega _{k}a_{k}^{+}a_{k}-\sum_{k}\frac{g_{k}^{2}}{4\omega
_{k}}\xi _{k}^{A}(2-\xi _{k}^{A}),
\end{equation}%
\begin{equation}
H_{0B}^{^{\prime }}=-\frac{1}{2}\eta ^{B}\Delta _{B}\sigma
_{z}^{B}+\sum_{k}\nu _{k}b_{k}^{+}b_{k}-\sum_{k}\frac{f_{k}^{2}}{4\nu _{k}}%
\xi _{k}^{B}(2-\xi _{k}^{B}).
\end{equation}%
As the same style, $H_{1}^{^{\prime }}=H_{1A}^{^{\prime }}+H_{1B}^{^{\prime
}}$ and $H_{2}^{^{\prime }}=H_{2A}^{^{\prime }}+H_{2B}^{^{\prime }}$, where%
\begin{equation}
H_{1A}^{^{\prime }}=\frac{1}{2}\sum_{k}\eta ^{A}\Delta _{A}\frac{g_{k}\xi
_{k}^{A}}{\omega _{k}}(a_{k}^{+}\sigma _{-}^{A}+a_{k}^{-}\sigma _{+}^{A}),
\end{equation}%
\begin{eqnarray}
H_{2A}^{^{\prime }} &=&-\frac{1}{2}\Delta \sigma _{x}\left[ \cosh (\sum_{k}%
\frac{g_{k}}{\omega _{k}}\xi _{k}(b_{k}^{+}-b_{k}))-\eta \right]  \notag \\
&&-i\frac{\Delta }{2}\sigma _{y}\left[ \sinh (\sum_{k}\frac{g_{k}}{\omega
_{k}}\xi _{k}(b_{k}^{+}-b_{k}))-\eta \sum_{k}\frac{g_{k}}{\omega _{k}}\xi
_{k}(b_{k}^{+}-b_{k})\right] ,
\end{eqnarray}%
with%
\begin{equation}
\eta ^{A}=\exp [-\sum_{k}\frac{g_{k}^{2}}{2\omega _{k}^{2}}(\xi
_{k}^{A})^{2}],\eta ^{B}=\exp [-\sum_{k}\frac{f_{k}^{2}}{2\nu _{k}^{2}}(\xi
_{k}^{B})^{2}]
\end{equation}%
\begin{equation}
\xi _{k}^{A}=\frac{\omega _{k}}{\omega _{k}+\eta ^{A}\Delta _{A}},\xi
_{k}^{B}=\frac{\nu _{k}}{\nu _{k}+\eta ^{B}\Delta _{B}}.
\end{equation}%
Here $\sigma _{\pm }^{A}=\sigma _{x}^{A}\mp \sigma _{y}^{A},$ $%
H_{0}^{^{\prime }}$ is the Hamiltonian of the noninteracting qubits and
environment, $H_{1}^{^{\prime }}$ and $H_{2}^{^{\prime }}$ are the
interaction Hamiltonian in increasing order of the qubit-environment
coupling strength $g_{k}$ and $f_{k}$. Comparing $H_{1}$ to $H_{1}^{^{\prime
}},$ the term of $H_{1}$ is replaced by the similar rotating-wave
approximation term in $H_{1}^{^{\prime }},$ while the qubit-environment
coupling strength $g_{k}$ in $H_{1}$ is replaced by $g_{k}\eta ^{A}\Delta
_{A}/(\omega _{k}+$ $\eta ^{A}\Delta _{A})$ in $H_{1}^{^{\prime }}.$ As we
seen, $g_{k}\eta ^{A}\Delta _{A}/(\omega _{k}+$ $\eta ^{A}\Delta
_{A})<g_{k}, $ that is to say, the counter-rotating terms decrease the
coupling strength with the environment.

Approximately to the order of $g_{k}^{2}$ and $f_{k}^{2}$, we write the
total Hamiltonian as $H^{^{\prime }}=H_{0}^{^{\prime }}+H_{1}^{^{\prime }}$.
In the interaction picture,
\begin{eqnarray}
V_{I}^{^{\prime }}(t) &=&\sum_{k}\eta ^{A}\Delta _{A}\frac{g_{k}\xi _{k}^{A}%
}{\omega _{k}}a_{k}^{+}\sigma _{-}^{A}\exp \left[ i(\omega _{k}-\eta
^{A}\Delta _{A})t\right] \\
&&+\sum_{k}\eta ^{A}\Delta _{A}\frac{g_{k}\xi _{k}^{A}}{\omega _{k}}%
a_{k}\sigma _{+}^{A}\exp \left[ -i(\omega _{k}-\eta ^{A}\Delta _{A})t\right]
\notag \\
&&+\sum_{k}\eta ^{B}\Delta _{B}\frac{f_{k}\xi _{k}^{B}}{\nu _{k}}%
b_{k}^{+}\sigma _{-}^{B}\exp \left[ i(\nu _{k}-\eta ^{B}\Delta _{B})t\right]
\notag \\
&&+\sum_{k}\eta ^{B}\Delta _{B}\frac{f_{k}\xi _{k}^{B}}{\nu _{k}}b_{k}\sigma
_{+}^{B}\exp \left[ -i(\nu _{k}-\eta ^{B}\Delta _{B})t\right] .  \notag
\end{eqnarray}%
We consider in general a system denoted by S interacting with a reservoir or
environment denoted by R. The combined density operator is denoted by $\rho
_{SR}$. The reduced density operator for the system $\rho _{S}$ is obtained
by taking a trace over the reservoir coordinates, i.e., $\rho
_{S}=Tr_{R}(\rho _{SR})$. The equation of motion for $\rho _{SR}$ is given
by
\begin{equation}
\frac{d}{dt}\rho _{SR}(t)=-i[V_{I}(t),\rho _{SR}].
\end{equation}%
After S transformation,%
\begin{equation}
\frac{d}{dt}\rho _{SR}^{^{\prime }}(t)=-i[V_{I}^{^{\prime }}(t),\rho
_{SR}^{^{\prime }}].
\end{equation}%
This equation can be formally integrated, and we obtain%
\begin{equation}
\rho _{SR}^{^{\prime }}(t)=\rho _{SR}^{^{\prime
}}(t_{i})-i\int\limits_{t_{i}}^{t}[V_{I}^{^{\prime }}(t^{^{\prime }}),\rho
_{SR}^{^{\prime }}(t^{^{\prime }})]dt^{^{\prime }}.
\end{equation}%
Here $t_{i}$ is an initial time when the interaction starts, supposing $%
t_{i}=0$. On substituting $\rho _{SR}^{^{\prime }}(t)$ into Eq.(18), we find
the equation of motion%
\begin{equation}
\frac{d}{dt}\rho _{SR}^{^{\prime }}(t)=-i[V_{I}^{^{\prime }}(t),\rho
_{SR}^{^{\prime }}(0)]-\int\limits_{0}^{t}[V_{I}^{^{\prime
}}(t),[V_{I}^{^{\prime }}(t^{^{\prime }}),\rho _{SR}^{^{\prime
}}(t^{^{\prime }})]]dt^{^{\prime }}.
\end{equation}%
We now employ the Born approximation\cite{F1,C1,C2} in which the interaction
between the qubit system and the environment is suppose to be weak, and
there is no back reaction effect of the qubits on the enviornment. In this
approximation, the state of the environment does not change in time, and we
can write the density operator $\rho _{SR}^{^{\prime }}(t)$ as $\rho
_{SR}^{^{\prime }}(t)=\rho _{S}^{^{\prime }}(t)\rho _{R}^{^{\prime }}(0).$
Under this approximation, Eq.(19) simplifies to
\begin{equation}
\frac{d}{dt}\rho _{S}^{^{\prime }}(t)\rho _{R}^{^{\prime
}}(0)=-i[V_{I}^{^{\prime }}(t),\rho _{S}^{^{\prime }}(0)\rho _{R}^{^{\prime
}}(0)]-\int\limits_{0}^{t}[V_{I}^{^{\prime }}(t),[V_{I}^{^{\prime
}}(t^{^{\prime }}),\rho _{S}^{^{\prime }}(t^{^{\prime }})\rho
_{R}(0)]]dt^{^{\prime }}.
\end{equation}%
Substituting $V_{I}^{^{\prime }}(t)$ into Eq.(20) and assuming that the
environment modes is in thermalization, the $Tr_{R}$ are given by:
\begin{eqnarray}
Tr_{R}[b_{k}^{+}b_{k}\rho _{R}] &=&Tr_{R}[b_{k}\rho _{R}b_{k}^{+}]=n_{k}, \\
Tr_{R}[b_{k}b_{k}^{+}\rho _{R}] &=&Tr_{R}[b_{k}^{+}\rho _{R}b_{k}]=n_{k}+1.
\end{eqnarray}%
Then,%
\begin{eqnarray}
&&\frac{d}{dt}\rho _{S}^{^{\prime }}(t) \\
&=&-\int\limits_{0}^{t}\sum_{k}(\eta ^{A}\Delta _{A}\frac{g_{k}\xi _{k}^{A}}{%
\omega _{k}})^{2}n_{k}^{A}[\sigma _{-}^{A}\sigma _{+}^{A}\rho _{S}^{^{\prime
I}}(t^{^{\prime }})-\sigma _{+}^{A}\rho _{S}^{^{\prime I}}(t^{^{\prime
}})\sigma _{-}^{A}]\exp \left[ i(\omega _{k}-\eta ^{A}\Delta
_{A})(t-t^{^{\prime }})\right] dt^{^{\prime }}  \notag \\
&&-\int\limits_{0}^{t}\sum_{k}(\eta ^{A}\Delta _{A}\frac{g_{k}\xi _{k}^{A}}{%
\omega _{k}})^{2}(n_{k}^{A}+1)[\rho _{S}^{^{\prime I}}(t^{^{\prime }})\sigma
_{+}^{A}\sigma _{-}^{A}-\sigma _{-}^{A}\rho _{S}^{^{\prime I}}(t^{^{\prime
}})\sigma _{+}^{A}]\exp \left[ i(\omega _{k}-\eta ^{A}\Delta
_{A})(t-t^{^{\prime }})\right] dt^{^{\prime }}  \notag \\
&&-\int\limits_{0}^{t}\sum_{k}(\eta ^{A}\Delta _{A}\frac{g_{k}\xi _{k}^{A}}{%
\omega _{k}})^{2}n_{k}^{A}[\rho _{S}^{^{\prime I}}(t^{^{\prime }})\sigma
_{-}^{A}\sigma _{+}^{A}-\sigma _{+}^{A}\rho _{S}^{^{\prime I}}(t^{^{\prime
}})\sigma _{-}^{A}]\exp \left[ -i(\omega _{k}-\eta ^{A}\Delta
_{A})(t-t^{^{\prime }})\right] dt^{^{\prime }}  \notag \\
&&-\int\limits_{0}^{t}\sum_{k}(\eta ^{A}\Delta _{A}\frac{g_{k}\xi _{k}^{A}}{%
\omega _{k}})^{2}(n_{k}^{A}+1)[\sigma _{+}^{A}\sigma _{-}^{A}\rho
_{S}^{^{\prime I}}(t^{^{\prime }})-\sigma _{-}^{A}\rho _{S}^{^{\prime
I}}(t^{^{\prime }})\sigma _{+}^{A}]\exp \left[ -i(\omega _{k}-\eta
^{A}\Delta _{A})(t-t^{^{\prime }})\right] dt^{^{\prime }}  \notag \\
&&-\int\limits_{0}^{t}\sum_{k}(\eta ^{B}\Delta _{B}\frac{f_{k}\xi _{k}^{B}}{%
\nu _{k}})^{2}n_{k}^{B}[\sigma _{-}^{B}\sigma _{+}^{B}\rho _{S}^{^{\prime
I}}(t^{^{\prime }})-\sigma _{+}^{B}\rho _{S}^{^{\prime I}}(t^{^{\prime
}})\sigma _{-}^{B}]\exp \left[ i(\nu _{k}-\eta ^{B}\Delta
_{B})(t-t^{^{\prime }})\right] dt^{^{\prime }}  \notag \\
&&-\int\limits_{0}^{t}\sum_{k}(\eta ^{B}\Delta _{B}\frac{f_{k}\xi _{k}^{B}}{%
\nu _{k}})^{2}(n_{k}^{B}+1)[\rho _{S}^{^{\prime I}}(t^{^{\prime }})\sigma
_{+}^{B}\sigma _{-}^{B}-\sigma _{-}^{B}\rho _{S}^{^{\prime I}}(t^{^{\prime
}})\sigma _{+}^{B}]\exp \left[ i(\nu _{k}-\eta ^{B}\Delta
_{B})(t-t^{^{\prime }})\right] dt^{^{\prime }}  \notag \\
&&-\int\limits_{0}^{t}\sum_{k}(\eta ^{B}\Delta _{B}\frac{f_{k}\xi _{k}^{B}}{%
\nu _{k}})^{2}n_{k}^{B}[\rho _{S}^{^{\prime I}}(t^{^{\prime }})\sigma
_{-}^{B}\sigma _{+}^{B}-\sigma _{+}^{B}\rho _{S}^{^{\prime I}}(t^{^{\prime
}})\sigma _{-}^{B}]\exp \left[ -i(\nu _{k}-\eta ^{B}\Delta
_{B})(t-t^{^{\prime }})\right] dt^{^{\prime }}  \notag \\
&&-\int\limits_{0}^{t}\sum_{k}(\eta ^{B}\Delta _{B}\frac{f_{k}\xi _{k}^{B}}{%
\nu _{k}})^{2}(n_{k}^{B}+1)[\sigma _{+}^{B}\sigma _{-}^{B}\rho
_{S}^{^{\prime I}}(t^{^{\prime }})-\sigma _{-}^{B}\rho _{S}^{^{\prime
I}}(t^{^{\prime }})\sigma _{+}^{B}]\exp \left[ -i(\nu _{k}-\eta ^{B}\Delta
_{B})(t-t^{^{\prime }})\right] dt^{^{\prime }}.  \notag
\end{eqnarray}%
In this equation, the $n_{k}$ and $n_{k}+1$ term on the right hand side
describe, respectively, decay and excitation process, with rate which depend
on the temperature, here parameterized by $n_{k}$, the average thermal
excitation of the reservoir. In this work, we study the limit of zero
temperature, $n_{k}=0,$ that is to say only the spontaneous decay term
survives leading to purely dissipative process.

The matrix equation is solved in the representation spanned by the standard
two-qubit product states basis $|1\rangle =|\uparrow \uparrow \rangle
,|2\rangle =|\uparrow \downarrow \rangle ,|3\rangle =|\downarrow \uparrow
\rangle ,|4\rangle =|\downarrow \downarrow \rangle .$ After Laplace
transformation and convolution theorem, the master equation of the system of
two qubits can be obtained as follow\cite{G1}:%
\begin{eqnarray}
P\overline{\rho _{S}^{^{\prime }}(P)}-\rho _{S}^{^{\prime }}(0) &=&-\sum_{k}%
\frac{(\eta ^{A}\Delta _{A}\frac{g_{k}\xi _{k}^{A}}{\omega _{k}})^{2}}{%
P-i(\omega _{k}-\eta ^{A}\Delta _{A})}[\overline{\rho _{S}^{^{\prime }}(P)}%
\sigma _{+}^{A}\sigma _{-}^{A}-\sigma _{-}^{A}\overline{\rho _{S}^{^{\prime
}}(P)}\sigma _{+}^{A}] \\
&&-\sum_{k}\frac{(\eta ^{B}\Delta _{B}\frac{g_{k}\xi _{k}^{B}}{\nu _{k}})^{2}%
}{P-i(\nu _{k}-\eta ^{B}\Delta _{B})}[\overline{\rho _{S}^{^{\prime }}(P)}%
\sigma _{+}^{B}\sigma _{-}^{B}-\sigma _{-}^{B}\overline{\rho _{S}^{^{\prime
}}(P)}\sigma _{+}^{B}]  \notag \\
&&-\sum_{k}\frac{(\eta ^{A}\Delta _{A}\frac{g_{k}\xi _{k}^{A}}{\omega _{k}}%
)^{2}}{P+i(\omega _{k}-\eta ^{A}\Delta _{A})}[\sigma _{+}^{A}\sigma _{-}^{A}%
\overline{\rho _{S}^{^{\prime }}(P)}-\sigma _{-}^{A}\overline{\rho
_{S}^{^{\prime }}(P)}\sigma _{+}^{A}]  \notag \\
&&-\sum_{k}\frac{(\eta ^{B}\Delta _{B}\frac{g_{k}\xi _{k}^{B}}{\nu _{k}})^{2}%
}{P+i(\nu _{k}-\eta ^{B}\Delta _{B})}[\sigma _{+}^{B}\sigma _{-}^{B}%
\overline{\rho _{S}^{^{\prime }}(P)}-\sigma _{-}^{B}\overline{\rho
_{S}^{^{\prime }}(P)}\sigma _{+}^{B}].  \notag
\end{eqnarray}%
Denote the summation of the environment degree of freedom $A_{+}=\sum_{k}%
\frac{(\eta ^{A}\Delta _{A}\frac{g_{k}\xi _{k}^{A}}{\omega _{k}})^{2}}{%
P+i(\omega _{k}-\eta ^{A}\Delta _{A})}$, $A_{-}=\sum_{k}\frac{(\eta
^{A}\Delta _{A}\frac{g_{k}\xi _{k}^{A}}{\omega _{k}})^{2}}{P-i(\omega
_{k}-\eta ^{A}\Delta _{A})}$, $B_{+}=\sum_{k}\frac{(\eta ^{B}\Delta _{B}%
\frac{g_{k}\xi _{k}^{B}}{\nu _{k}})^{2}}{P+i(\nu _{k}-\eta ^{B}\Delta _{B})}$%
and $B_{-}=\sum_{k}\frac{(\eta ^{B}\Delta _{B}\frac{g_{k}\xi _{k}^{B}}{\nu
_{k}})^{2}}{P-i(\nu _{k}-\eta ^{B}\Delta _{B})}.$ The decay rate is
dependent on the process, as seen from $A_{+},$ $A_{-},$ $B_{+}$ and $B_{-},$
instead of constant for all process in Markovian approximation. We shall
therefore focus on the precise time scales of every decay process.

According to the Kronecker product\ property and technique to Lyapunov
matrix equation in matrix theory, expand matrix into vector along row of the
matrix from two sides of master equation,%
\begin{eqnarray}
&&\bigg\{PI_{16\times 16}+[A_{-}I_{4\times 4}\otimes (\sigma _{+}^{A}\sigma
_{-}^{A}\otimes I_{2\times 2})^{T}+B_{-}I_{4\times 4}\otimes (I_{2\times
2}\otimes \sigma _{+}^{B}\sigma _{-}^{B})^{T}] \\
&&-[A_{-}(\sigma _{-}^{A}\otimes I_{2\times 2})\otimes (\sigma
_{+}^{A}\otimes I_{2\times 2})^{T}+B_{-}(I_{2\times 2}\otimes \sigma
_{-}^{B})\otimes (I_{2\times 2}\otimes \sigma _{+}^{B})^{T}]  \notag \\
&&-[A_{+}(\sigma _{-}^{A}\otimes I_{2\times 2})\otimes (\sigma
_{+}^{A}\otimes I_{2\times 2})^{T}+B_{+}(I_{2\times 2}\otimes \sigma
_{-}^{B})\otimes (I_{2\times 2}\otimes \sigma _{+}^{B})^{T}]  \notag \\
&&+[A_{+}(\sigma _{+}^{A}\sigma _{-}^{A}\otimes I_{2\times 2})\otimes
I_{4\times 4}+B_{-}(I_{2\times 2}\otimes \sigma _{+}^{B}\sigma
_{-}^{B})\otimes I_{4\times 4}]\bigg\}Vec[\overline{\rho _{S}^{^{\prime }}(P)%
}]=Vec[\rho _{S}^{^{\prime }}(0)].  \notag
\end{eqnarray}%
The $4\times 4$ matrix equation transformed into $16\times 16$ matrix
equation with the form%
\begin{equation}
U(P)_{16\times 16}Vec[\overline{\rho _{S}^{^{\prime }}(P)}]=Vec[\rho
_{S}^{^{\prime }}(0)]
\end{equation}%
where $Vec[\overline{\rho _{S}^{^{\prime }}(P)}]$ is the vector of row
expanding of matrix $\overline{\rho _{S}^{^{\prime }}(P)}$. The solution
formally is%
\begin{equation}
Vec[\overline{\rho _{S}^{^{\prime }}(P)}]=U(P)_{16\times 16}^{-1}Vec[\rho
_{S}^{^{\prime }}(0)].
\end{equation}%
Inverse Laplace transformation to time parameter space,%
\begin{equation}
\mathscr{L}^{-1}Vec[\overline{\rho _{S}^{^{\prime }}(P)}]=\mathscr{L}%
^{-1}U(P)_{16\times 16}^{-1}Vec[\rho _{S}^{^{\prime }}(0)].
\end{equation}%
i.e.%
\begin{equation}
Vec[\rho _{S}^{^{\prime I}}(t)]=\mathscr{L}^{-1}U(P)_{16\times
16}^{-1}Vec[[\rho _{S}^{^{\prime }}(0)]].
\end{equation}%
$\mathscr{L}^{-1}U(P)_{16\times 16}^{-1}$ can be obtained (see Appendix).

Compared with Markovian approximation, decoherence rates $\gamma (\omega )$
in our results becomes frequency dependent. Due to entanglement and
environment interaction together, the decay rate for variety process are
different, some increase slower, some increase faster, i.e. the two bathes
has indirect interaction through the two entangled qubits. That is more
general and physical.

Therefore, the reduced density matrix $\rho _{S}^{^{\prime }}(t)$ in the
Schrodinger picture is obtained $\rho _{S}^{^{\prime }}(t)=\exp
(-iH_{0}^{^{\prime }}t)\rho _{S}^{^{\prime I}}(t)\exp (iH_{0}^{^{\prime }}t)$%
, the matrix form is%
\begin{eqnarray}
\rho _{S}^{^{\prime }}(t) &=&\left[ \left(
\begin{array}{cc}
\exp (i\frac{\eta ^{A}\Delta _{A}}{2}t) & 0 \\
0 & \exp (-i\frac{\eta ^{A}\Delta _{A}}{2}t)%
\end{array}%
\right) \otimes \left(
\begin{array}{cc}
\exp (i\frac{\eta ^{B}\Delta _{B}}{2}t) & 0 \\
0 & \exp (-i\frac{\eta ^{B}\Delta _{B}}{2}t)%
\end{array}%
\right) \right] \\
&&\rho _{S}^{^{\prime I}}(t)\left[ \left(
\begin{array}{cc}
\exp (-i\frac{\eta ^{A}\Delta _{A}}{2}t) & 0 \\
0 & \exp (i\frac{\eta ^{A}\Delta _{A}}{2}t)%
\end{array}%
\right) \otimes \left(
\begin{array}{cc}
\exp (-i\frac{\eta ^{B}\Delta _{B}}{2}t) & 0 \\
0 & \exp (i\frac{\eta ^{B}\Delta _{B}}{2}t)%
\end{array}%
\right) \right] .  \notag
\end{eqnarray}%
Transform $\rho _{S}^{^{\prime }}(t)$ into $\rho _{S}(t)$ through $\rho
_{S}(t)=Tr_{R}[\exp (-S)\rho _{S}^{^{\prime }}(t)\rho _{R}(0)\exp (S)],$
denoting $X_{A}=\sum_{k}\frac{g_{k}}{2\omega _{k}}\xi
_{k}^{A}(a_{k}^{+}-a_{k}),X_{B}=\sum_{k}\frac{f_{k}}{2\nu _{k}}\xi
_{k}^{B}(b_{k}^{+}-b_{k}),$ so
\begin{eqnarray}
\rho _{S}(t) &=&Tr_{R}[(\cosh X_{A}-\sinh X_{A}\sigma _{x}^{A})\otimes
(\cosh X_{B}-\sinh X_{B}\sigma _{x}^{B})\rho _{S}^{^{\prime }}(t)\rho _{R} \\
&&(\cosh X_{A}+\sinh X_{A}\sigma _{x}^{A})\otimes (\cosh X_{B}+\sinh
X_{B}\sigma _{x}^{B})  \notag \\
&=&\frac{1+\eta ^{A}}{2}\frac{1+\eta ^{B}}{2}\rho _{S}^{^{\prime }}(t)+\frac{%
1+\eta ^{A}}{2}\frac{1-\eta ^{B}}{2}(I_{2\times 2}\otimes \sigma
_{x}^{B})\rho _{S}^{^{\prime }}(t)(I_{2\times 2}\otimes \sigma _{x}^{B})
\notag \\
&&+\frac{1-\eta ^{A}}{2}\frac{1+\eta ^{B}}{2}(\sigma _{x}^{A}\otimes
I_{2\times 2})\rho _{S}^{^{\prime }}(t)(\sigma _{x}^{A}\otimes I_{2\times 2})
\notag \\
&&+\frac{1-\eta ^{A}}{2}\frac{1-\eta ^{B}}{2}(\sigma _{x}^{A}\otimes \sigma
_{x}^{B})\rho _{S}^{^{\prime }}(t)(\sigma _{x}^{A}\otimes \sigma _{x}^{B}).
\notag
\end{eqnarray}%
Until now, we obtain the reduced density matrix in all kinds of initial
state.

Although a general solution to this problem, for arbitrary system dynamics
and decoherence mechanisms is still out of reach, out technical machinery,
developed in the previous section allows to treat arguably all situations
encountered in typical state of the art experiments, as in the quantum
optics and condensed matter.

\section{The result and discussion}

\label{discussion}

We assume that at t=0, the two qubits and environment are described by the
product state $\exp (-S)\rho _{S}(0)\rho _{R}(0)\exp (S)=\Psi _{AB}\otimes
|0\rangle _{A}|0\rangle _{B}$, where $\Psi _{AB}$ is the entangled initial
state of the two qubits and $|0\rangle _{A}|0\rangle _{B}$ is the vacuum
state of two environments. Let us assume that the initial density matrix is
only practically coherence of a familiar type (one of the atoms is excited,
but it is not certain which one). This is easily expressed in the following
form\cite{H3,C4}%
\begin{equation}
\rho _{S}^{^{\prime }}(0)=\frac{1}{3}\left(
\begin{array}{cccc}
d & 0 & 0 & 0 \\
0 & c & z & 0 \\
0 & z^{\ast } & b & 0 \\
0 & 0 & 0 & a%
\end{array}%
\right) .
\end{equation}%
where the factor 1/3 is for notational convenience. In order to compare with
previous results, consider an important class of mixed state with single
parameter $a$ satisfying initially $a\geq 0,$ $d=1-a,$ and $b=c=z=1.$ We
will use Wootter's concurrence to quantify the degree of entanglement\cite%
{H1,H2}. Let $\rho $ be density matrix of the pair of qubits expressed in
the standard basis. The concurrence may be calculated explicitly from the
density matrix $\rho $ for qubits A and B: $C=\max (0,\sqrt{\lambda _{1}}-%
\sqrt{\lambda _{2}}-\sqrt{\lambda _{3}}-\sqrt{\lambda _{4}}),$ where the
quantities $\lambda _{i}$ are the eigenvalues of the matrix $M$: $M=\rho
(\sigma _{y}^{A}\otimes \sigma _{y}^{B})\rho ^{\ast }(\sigma _{y}^{A}\otimes
\sigma _{y}^{B}),$ arranged in decreasing order. Here $\rho ^{\ast }$
denotes the complex conjugation of $\rho $ in the standard basis. It can be
shown that the concurrence varies from 0 for a disentangled state to $C=1$
for a maximally entangled state.

Firstly consider very weak qubit-environment interaction, $\alpha
_{A}=\alpha _{B}=0.01$, which is larger a bit than the fine-structure
constant $1/137.$ Here and in the following, energies $\Delta _{A},$ and $%
\Delta _{B}$ are expressed in units of $\omega _{c},$ times in units of $%
\omega _{c}^{-1}.$ We assume $\Delta _{A}=0.2,$ $\Delta _{B}=0.4.$ In Fig.1,
the time evolution of the concurrence for various values of the parameter $a$
is shown. The figure shows that for all $a$ values almost between 0.3 and 1,
concurrence decays is completed in a finite-time, which is the effect of
"entanglement sudden death" \cite{B1,C3}, but for smaller $a$'s the time for
completed decay is infinite, which is consistent with Ref.15 and 19. The
result indicated that in the weak dissipation environment, such as the
all-optical setup in Ref.15, the Markovian approximation and rotating-wave
approximation are available. When the coupling constant to the enviornment
is near to or less than fine-structure constant, we see that the quantum
dissipation of the vacuum environment is not sufficient to completely
destroy the entanglement in a finite time in some situations. The sudden
death of entanglement results from the decays of the mixed double excitation
state component. With increasing of the mixed double excitation state
component, $a$ value, concurrence change from exponential decay to sudden
death. The entanglement has another unusual relaxation property: different
entangled states, corresponding to different values of $a,$ with the same
initial degrees of entanglement may evolve with different route, some
showing entanglement sudden death, some not, some decay faster, some slower.
That is to say, we can prepare certain initial state to prolong entanglement
time.

Next, consider large qubit-environment interaction, $\alpha _{A}=\alpha
_{B}=0.05,$ the other parameters and initial entangled state are same with
Fig.1. The time evolution of the concurrence through the entire range of
different $a$ values is plotted in Fig.2. As we shown, concurrence actually
goes abruptly to zero in a finite time and remains zero thereafter. That is
to say the entanglement sudden-death always happens. In the first example
above, we have shown that the entanglement can last for infinite period in
the vacuum reservoir for some initial entangle state. However, in Fig.2 the
sudden death of entanglement always happens no matter which entangled state
the qubit are initially in. That is also shown that the disentanglement
dynamics varies with the coupling strength with the enviornment or the
rotating-wave approximation and Markovian approximation is unavailable, when
the coupling to the environment is much larger than the fine-structure
constant. Fig.3 shows concurrence for $\alpha _{A}=\alpha _{B}=0.1,$ under
the same initial condition. It is observed that in the same initial state,
the death time decreases as the increasing of the strength of
qubit-environment interaction.

\section{Conclusion}

\label{conclusion}

In this paper, we consider two two-level qubits that are spatially separated
from each other and independently coupled to local vacuum environments. We
investigated the dynamics evolution of entanglement between the qubits. We
show that, for a certain class of two-qubit entangled state, the
entanglement measured by concurrence can change from exponential decay to
sudden death with increasing of the mixed double excitation state component
in the case of weak coupling with environment. Increasing coupling strength,
the entanglement sudden-death always happens no matter which entangled state
the qubit are initially in. The exponential decay of entanglement is a very
special result to the weak dissipation vacuum reservoir. The entangle sudden
death time in our result is obtained from the physical parameter: coupling
constant $\alpha $, energy splitting $\Delta _{A},$ $\Delta _{B}$ and
cut-off frequency $\omega _{c}$. Finally, we hope that this work will
stimulate more experimental and theoretical works in quantum information and
computation for quantum optical control.

\vskip 0.5cm

{\noindent {\large \textbf{Acknowledgement}}}

\bigskip This work was supported by the China National Natural Science
Foundation (Grants No. 10474062 and No. 90503007).

\section*{Appendix}

\setcounter{equation}{0} \renewcommand{\theequation}{A\arabic{equation}}

In this Appendix, we give details of how to inverse Laplace transformation
to time parameter space. $U(P)_{16\times 16}^{-1}$ is composed by the matrix
element$:$ $\frac{1}{P},$ $\frac{1}{P+A_{-}},$ $\frac{1}{P+A_{+}},$ $\frac{1%
}{P+A_{-}+A_{+}},\frac{1}{P+A_{-}+B_{+}},\frac{1}{P+A_{-}+A_{+}+B_{+}},\frac{%
1}{P+A_{-}+A_{+}+B_{+}+B_{-}}$ \textit{etc}. Then $\mathscr{L}%
^{-1}U(P)_{16\times 16}^{-1}$ is inverse every matrix element. As we know, $%
\mathscr{L}^{-1}\frac{1}{P}=1.$ Solve $\mathscr{L}^{-1}\frac{1}{P+A_{-}}$
\textit{etc.} through the following method.
\begin{equation}
\mathscr{L}^{-1}\frac{1}{P+A_{-}}=\frac{1}{2\pi i}\int\limits_{\sigma
-i\infty }^{\sigma +i\infty }\frac{\exp (Pt)}{P+\sum_{k}\frac{(\eta
^{A}\Delta _{A}\frac{g_{k}\xi _{k}^{A}}{\omega _{k}})^{2}}{P-i(\omega
_{k}-\eta ^{A}\Delta _{A})}}dP
\end{equation}%
Then Changing $P$ to $i\omega +0^{+},$\cite{I1}
\begin{equation}
\frac{1}{2\pi i}\int\limits_{\sigma -i\infty }^{\sigma +i\infty }\frac{\exp
(Pt)}{P+\sum_{k}\frac{(\eta ^{A}\Delta _{A}\frac{g_{k}\xi _{k}^{A}}{\omega
_{k}})^{2}}{P-i(\omega _{k}-\eta ^{A}\Delta _{A})}}dP=\frac{1}{2\pi i}%
\int\limits_{-\infty }^{+\infty }\frac{\exp (i\omega t+0^{+})}{\omega
-\sum_{k}\frac{(\eta ^{A}\Delta _{A}\frac{g_{k}\xi _{k}^{A}}{\omega _{k}}%
)^{2}}{(\omega +\eta ^{A}\Delta _{A})-\omega _{k}-i0^{+}}}d\omega .
\end{equation}%
Denote $R(\omega )$ and $\gamma (\omega )$ as the real and imaginary parts
of $\sum_{k}(\eta ^{A}\Delta _{A}\frac{g_{k}\xi _{k}^{A}}{\omega _{k}}%
)^{2}/(\omega -\omega _{k}-i0^{+}),$

\begin{eqnarray}
R(\omega ) &=&\sum_{k}\wp \frac{(\eta ^{A}\Delta _{A}\frac{g_{k}\xi _{k}^{A}%
}{\omega _{k}})^{2}}{\omega -\omega _{k}}=(\eta \Delta )^{2}\wp
\int_{0}^{\infty }d\omega ^{^{\prime }}\frac{J(\omega ^{^{\prime }})}{%
(\omega -\omega ^{^{\prime }})(\omega ^{^{\prime }}+\eta \Delta )^{2}}
\notag \\
&=&-2\alpha \frac{(\eta \Delta )^{2}}{\omega +\eta \Delta }\left\{ \frac{%
\omega _{c}}{\omega _{c}+\eta \Delta }-\frac{\omega }{\omega +\eta \Delta }%
\ln \left[ \frac{\left\vert \omega \right\vert (\omega _{c}+\eta \Delta )}{%
\eta \Delta (\omega _{c}-\omega )}\right] \right\} ,
\end{eqnarray}%
and%
\begin{eqnarray}
\gamma (\omega ) &=&\pi \sum_{k}(\eta ^{A}\Delta _{A}\frac{g_{k}\xi _{k}^{A}%
}{\omega _{k}})^{2}\delta (\omega -\omega _{k})=\pi (\eta \Delta )^{2}\frac{%
J(\omega )}{(\omega +\eta \Delta )^{2}}  \notag \\
&=&2\alpha \pi \omega \frac{(\eta \Delta )^{2}}{(\omega +\eta \Delta )^{2}}.
\end{eqnarray}%
Where $\wp $ stands for Cauchy principal value.
\begin{eqnarray}
\mathscr{L}^{-1}\frac{1}{P+A_{-}} &=&\frac{1}{2\pi i}\int\limits_{-\infty
}^{+\infty }\frac{\exp (i\omega t+0^{+})}{\omega -R(\omega +\eta ^{A}\Delta
_{A})+i\gamma (\omega +\eta ^{A}\Delta _{A})}d\omega \\
&=&\exp [i\omega _{01}t-\gamma (\omega _{01}+\eta ^{A}\Delta _{A})t]  \notag
\end{eqnarray}%
where $\omega _{01}$ is the solution of equation $\omega -R(\omega +\eta
^{A}\Delta _{A})=0$ and is the Lamb shift due to the local interaction of
the qubit with the enviornment.

In the same way,
\begin{equation}
\mathscr{L}^{-1}\frac{1}{P+A_{+}}=\exp [-i\omega _{01}t-\gamma (\omega
_{01}+\eta ^{A}\Delta _{A})t].
\end{equation}%
It is clear that $\mathscr{L}^{-1}\frac{1}{P+A_{-}}$ conjugate with $%
\mathscr{L}^{-1}\frac{1}{P+A_{+}}.$
\begin{eqnarray}
\mathscr{L}^{-1}\frac{1}{P+A_{-}+A_{+}} &=&\frac{1}{2\pi i}%
\int\limits_{-\infty }^{+\infty }\frac{\exp (i\omega t+0^{+})}{\omega
+i\gamma (\eta ^{A}\Delta _{A})+i\gamma (\eta ^{A}\Delta _{A})}d\omega  \\
&=&\exp [-2\gamma (\eta ^{A}\Delta _{A})t],
\end{eqnarray}%
The decay for $\mathscr{L}^{-1}\frac{1}{P+A_{-}+A_{+}}$ accelerated (by a
factor of almost two) as compared to $\mathscr{L}^{-1}\frac{1}{P+A_{+}}$,
under the influence of zero temperature environment.\cite{C2}%
\begin{eqnarray}
&&\mathscr{L}^{-1}\frac{1}{P+A_{-}+B_{-}} \\
&=&\frac{1}{2\pi i}\int\limits_{-\infty }^{+\infty }\frac{\exp (i\omega
t+0^{+})}{\omega -R(\omega +\eta ^{A}\Delta _{A})-R(\omega +\eta ^{B}\Delta
_{B})+i\gamma (\omega +\eta ^{A}\Delta _{A})+i\gamma (\omega +\eta
^{B}\Delta _{B})}d\omega   \notag \\
&=&\exp [i\omega _{12}^{s}t-\gamma (\omega _{12}^{s}+\eta ^{A}\Delta
_{A})t-\gamma (\omega _{12}^{s}+\eta ^{B}\Delta _{B})t],  \notag
\end{eqnarray}%
where $\omega _{12}^{s}$ is the solution of $\omega -R(\omega +\eta
^{A}\Delta _{A})-R(\omega +\eta ^{B}\Delta _{B})=0$ and is the Lamb shift
due to the two environments indirect interaction, which is a nonlocal
effect. $\mathscr{L}^{-1}\frac{1}{P+A_{+}+B_{+}}$ conjugates with $%
\mathscr{L}^{-1}\frac{1}{P+A_{-}+B_{-}}.$%
\begin{eqnarray}
&&\mathscr{L}^{-1}\frac{1}{P+A_{+}+B_{-}} \\
&=&\frac{1}{2\pi i}\int\limits_{-\infty }^{+\infty }\frac{\exp (i\omega
t+0^{+})}{\omega -R(\eta ^{A}\Delta _{A}-\omega )-R(\omega +\eta ^{B}\Delta
_{B})+i\gamma (\eta ^{A}\Delta _{A}-\omega )+i\gamma (\omega +\eta
^{B}\Delta _{B})}d\omega   \notag \\
&=&\exp [i\omega _{12}^{a}t-\gamma (\eta ^{A}\Delta _{A}-\omega
_{12}^{a})t-\gamma (\omega _{12}^{a}+\eta ^{B}\Delta _{B})t],  \notag
\end{eqnarray}%
where $\omega _{12}^{a}$ is the solution of $\omega -R(\eta ^{A}\Delta
_{A}-\omega )-R(\omega +\eta ^{B}\Delta _{B})=0$ and is also the Lamb shift
due to the two environment indirect interaction. In the same way, $%
\mathscr{L}^{-1}\frac{1}{P+A_{-}+B_{+}}$ conjugates with $\mathscr{L}^{-1}%
\frac{1}{P+A_{+}+B_{-}}.$%
\begin{eqnarray}
&&\mathscr{L}^{-1}\frac{1}{P+A_{+}+A_{-}+B_{-}} \\
&=&\frac{1}{2\pi i}\int\limits_{-\infty }^{+\infty }\exp (i\omega
t+0^{+})/[\omega -R(\eta ^{A}\Delta _{A}-\omega )-R(\eta ^{A}\Delta
_{A}+\omega )-R(\omega +\eta ^{B}\Delta _{B})  \notag \\
&&+i\gamma (\eta ^{A}\Delta _{A}-\omega )+i\gamma (\eta ^{A}\Delta
_{A}+\omega )+i\gamma (\omega +\eta ^{B}\Delta _{B})]d\omega  \\
&=&\exp [i\omega _{31}t-\gamma (\eta ^{A}\Delta _{A}-\omega _{31})t-\gamma
(\eta ^{A}\Delta _{A}+\omega _{31})t-\gamma (\omega _{31}+\eta ^{B}\Delta
_{B})t],  \notag
\end{eqnarray}%
where $\omega _{31}$ is the solution of $\omega -R(\eta ^{A}\Delta
_{A}-\omega )-R(\eta ^{A}\Delta _{A}+\omega )-R(\omega +\eta ^{B}\Delta
_{B})=0$ and is another Lamb shift due to nonlocal interaction. $\mathscr{L}%
^{-1}\frac{1}{P+A_{+}+A_{-}+B_{+}}$ conjugates with $\mathscr{L}^{-1}\frac{1%
}{P+A_{+}+A_{-}+B_{-}}.$%
\begin{eqnarray}
&&\mathscr{L}^{-1}\frac{1}{P+A_{-}+B_{+}+B_{-}} \\
&=&\frac{1}{2\pi i}\int\limits_{-\infty }^{+\infty }\exp (i\omega
t+0^{+})/[\omega -R(\omega +\eta ^{A}\Delta _{A})-R(\eta ^{B}\Delta
_{B}-\omega )-R(\eta ^{B}\Delta _{B}+\omega )  \notag \\
&&+i\gamma (\omega +\eta ^{A}\Delta _{A})+i\gamma (\eta ^{B}\Delta
_{B}-\omega )+i\gamma (\eta ^{B}\Delta _{B}+\omega )]d\omega   \notag \\
&=&\exp [i\omega _{32}t-\gamma (\omega _{32}+\eta ^{A}\Delta _{A})t-\gamma
(\eta ^{B}\Delta _{B}-\omega _{32})t-\gamma (\eta ^{B}\Delta _{B}+\omega
_{32})t],  \notag
\end{eqnarray}%
where $\omega _{32}$ is the solution of $\omega -R(\omega +\eta ^{A}\Delta
_{A})-R(\eta ^{B}\Delta _{B}-\omega )-R(\eta ^{B}\Delta _{B}+\omega )=0$ and
is the Lamb shift due to the two environment indirect interaction, too. $%
\mathscr{L}^{-1}\frac{1}{P+A_{-}+B_{+}+B_{-}}$ conjugates with $\mathscr{L}%
^{-1}\frac{1}{P+A_{+}+B_{+}+B_{-}}.$%
\begin{eqnarray}
&&\mathscr{L}^{-1}\frac{1}{P+A_{+}+A_{-}+B_{+}+B_{-}} \\
&=&\frac{1}{2\pi i}\int\limits_{-\infty }^{+\infty }\frac{\exp (i\omega
t+0^{+})}{\omega +i\gamma (\eta ^{A}\Delta _{A}-\omega )+i\gamma (\eta
^{A}\Delta _{A}+\omega )+i\gamma (\omega +\eta ^{B}\Delta _{B})+i\gamma
(\eta ^{B}\Delta _{B}-\omega )}d\omega   \notag \\
&=&\exp [-2\gamma (\eta ^{A}\Delta _{A})t-2\gamma (\eta ^{B}\Delta _{B})t].
\notag
\end{eqnarray}

\bigskip

\bigskip

--------------------

\begin{description}
\item {\large Figure Caption}

%\begin{figure}
\vspace{0.3cm} %\centerline{\psfig{figure=fig1.eps,width=8.5cm}}
\vspace{0.3cm} %\caption

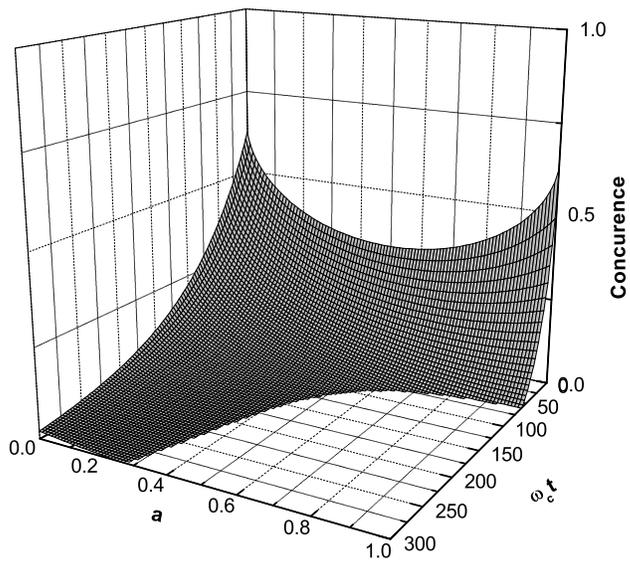
\item {Fig. 1. The entanglement decay via spontaneous emission of two
two-level qubits starting from the initially entangled state }${(1-a)/3}%
|\uparrow \uparrow \rangle \left\langle \uparrow \uparrow \right\vert
+a/3|\downarrow \downarrow \rangle \left\langle \downarrow \downarrow
\right\vert +1/3(|\downarrow \uparrow \rangle +|\uparrow \downarrow \rangle
)(\left\langle \uparrow \downarrow \right\vert +\left\langle \downarrow
\uparrow \right\vert )$ {with }${a}${\ between zero and }${1}${. }the
coupling constant of the environment and qubit $\alpha _{A}=\alpha
_{B}=0.01. $ Here and in the following figures energies $\Delta _{A}$ and $%
\Delta _{B}$ are expressed in units of $\omega _{c},$ times in units of $%
\omega _{c}^{-1}. $ We assume $\Delta _{A}=0.2,$ $\Delta _{B}=0.4.$

%\label{fig1}
%\end{figure}

%\begin{figure}
\vspace{0.3cm} %\centerline{\psfig{figure=fig2.eps,width=8.5cm}}
\vspace{0.3cm} %\caption

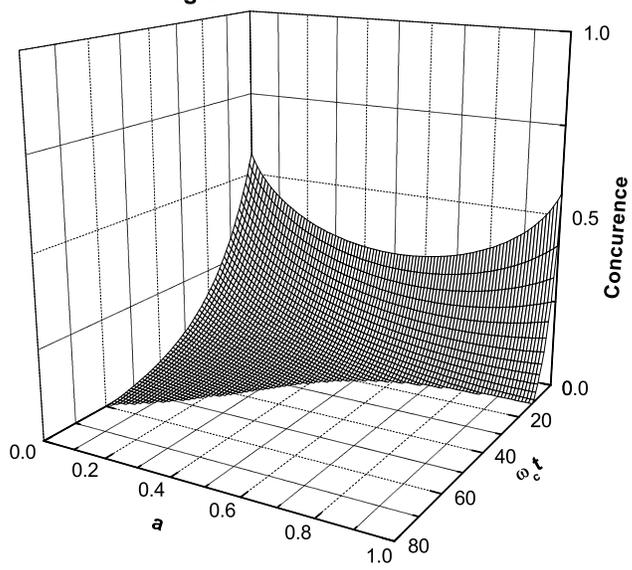
\item {Fig. 2. The entanglement decay via spontaneous emission of two
two-level qubits }the coupling constant of the environment and qubit $\alpha
_{A}=\alpha _{B}=0.05.$ the other parameter the same as Fig. 1.

%\label{fig1}
%\end{figure}

%\begin{figure}
\vspace{0.3cm} %\centerline{\psfig{figure=fig2.eps,width=8.5cm}}
\vspace{0.3cm} %\caption

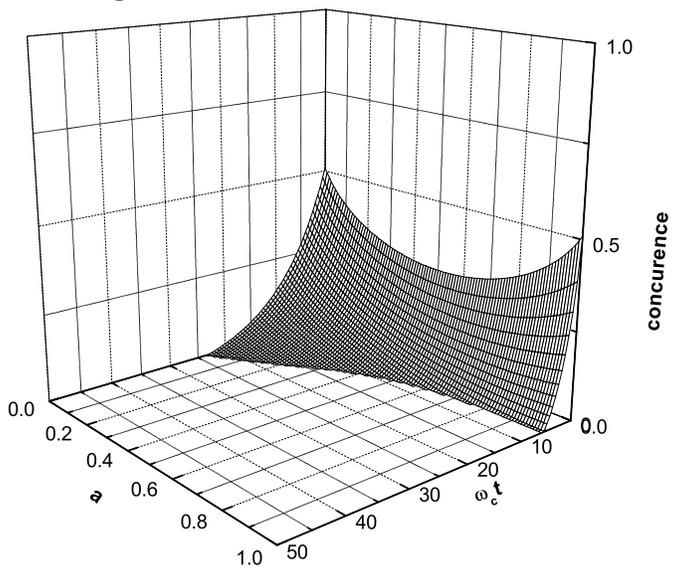
\item {Fig. 3. The entanglement decay via spontaneous emission of two
two-level qubits. }the coupling constant of the environment and qubit $%
\alpha _{A}=\alpha _{B}=0.1.$ the other parameter the same as Fig. 1.

%\label{fig1}
%\end{figure}
\end{description}

\end{document}